\documentclass[superscriptaddress,amsmath,amssymb,aps,prl,twocolumn,preprint, tightenlines]{revtex4-2}

\usepackage{graphicx}
\usepackage[hidelinks]{hyperref}

\usepackage{siunitx} 
\usepackage[version=4]{mhchem} 
\usepackage{tabularx} 

\usepackage{float}
\usepackage{subfigure}

\begin{document}


\title{
Isomeric excitation energy for $^{99}$In$^{m}$ from mass spectrometry reveals
constant trend next to doubly magic $^{100}$Sn
}

\author{L.~Nies}
\email{Lukas.Nies@cern.ch}
\affiliation{European Organization for Nuclear Research (CERN), 1211 Geneva 23, Switzerland}
\affiliation{Institut für Physik, Universität Greifswald, 17487 Greifswald, Germany}

\author{D.~Atanasov}
\altaffiliation{Present address: LP2i Bordeaux, UMR5797, Université de Bordeaux, CNRS, France}
\affiliation{European Organization for Nuclear Research (CERN), 1211 Geneva 23, Switzerland}

\author{M.~Athanasakis-Kaklamanakis}
\affiliation{European Organization for Nuclear Research (CERN), 1211 Geneva 23, Switzerland}
\affiliation{KU Leuven, Instituut voor Kern- en Stralingsfysica, B-3001 Leuven, Belgium}

\author{M.~Au}
\affiliation{European Organization for Nuclear Research (CERN), 1211 Geneva 23, Switzerland}
\affiliation{Johannes Gutenberg-Universit\"at Mainz, 55128 Mainz, Germany}

\author{K.~Blaum}
\affiliation{Max-Planck-Institut für Kernphysik, 69117 Heidelberg, Germany}

\author{J.~Dobaczewski}
\affiliation{School of Physics, Engineering and Technology, University of York, Heslington, York, UK}
\affiliation{Institute of Theoretical Physics, Faculty of Physics, University of Warsaw, Warsaw, Poland}

\author{B.~S.~Hu}
\affiliation{TRIUMF, Vancouver, British Columbia, Canada}

\author{J.~D.~Holt}
\affiliation{TRIUMF, Vancouver, British Columbia, Canada}
\affiliation{Department of Physics, McGill University, Montreal, Quebec, Canada}

\author{J.~Karthein}
\affiliation{Massachusetts Institute of Technology, Cambridge, MA-02139, USA}

\author{I.~Kulikov}
\affiliation{GSI Helmholtzzentrum für Schwerionenforschung GmbH, Darmstadt, Germany}

\author{Yu.~A.~Litvinov}
\affiliation{GSI Helmholtzzentrum für Schwerionenforschung GmbH, Darmstadt, Germany}
\affiliation{ExtreMe Matter Institute EMMI, GSI Helmholtzzentrum für Schwerionenforschung GmbH, 64291 Darmstadt, Germany}

\author{D.~Lunney}
\affiliation{Université Paris-Saclay, CNRS/IN2P3, IJCLab, 91405 Orsay, France}

\author{V.~Manea}
\affiliation{Université Paris-Saclay, CNRS/IN2P3, IJCLab, 91405 Orsay, France}

\author{T.~Miyagi}
\affiliation{Institut für Kernphysik, Technische Universität Darmstadt, 64289 Darmstadt, Germany}
\affiliation{ExtreMe Matter Institute EMMI, GSI Helmholtzzentrum für Schwerionenforschung GmbH, 64291 Darmstadt, Germany}
\affiliation{Max-Planck-Institut für Kernphysik, 69117 Heidelberg, Germany}

\author{M.~Mougeot}
\altaffiliation{University of Jyv\"{a}skyl\"{a}, Department of Physics, Accelerator laboratory, P.O. Box 35(YFL) FI-40014 University of Jyv\"{a}skyl\"{a}, Finland}
\affiliation{European Organization for Nuclear Research (CERN), 1211 Geneva 23, Switzerland}
\affiliation{Max-Planck-Institut für Kernphysik, 69117 Heidelberg, Germany}

\author{L.~Schweikhard}
\affiliation{Institut für Physik, Universität Greifswald, 17487 Greifswald, Germany}

\author{A.~Schwenk}
\affiliation{Institut für Kernphysik, Technische Universität Darmstadt, 64289 Darmstadt, Germany}
\affiliation{ExtreMe Matter Institute EMMI, GSI Helmholtzzentrum für Schwerionenforschung GmbH, 64291 Darmstadt, Germany}
\affiliation{Max-Planck-Institut für Kernphysik, 69117 Heidelberg, Germany}

\author{K.~Sieja}
\affiliation{IPHC, CNRS/IN2P3 et Université de Strasbourg, F-67037 Strasbourg, France}

\author{F.~Wienholtz}
\affiliation{Institut für Kernphysik, Technische Universität Darmstadt, 64289 Darmstadt, Germany}


\date{\today}

\begin{abstract}
	The excitation energy of the 1/2$^-$ isomer in \ce{^{99}In}{} at ${N=50}$ is measured to be $\SI{671\pm37}{\kilo\electronvolt}$ and the mass uncertainty of the 9/2$^+$ ground state is significantly reduced using the ISOLTRAP mass spectrometer at ISOLDE/CERN.
    The measurements exploit a major improvement in the resolution of the multi-reflection time-of-flight mass spectrometer.
    The results reveal an intriguing constancy of the $1/2^-$ isomer excitation energies in neutron-deficient indium that persists down to the $N = 50$ shell closure, even when all neutrons are removed from the valence shell.     
    This trend is used to test large-scale shell model, \textit{ab initio}, and density functional theory calculations. 
     The models have difficulties describing both the isomer excitation energies and ground-state electromagnetic moments along the indium chain.
\end{abstract}

\maketitle

Considerable experimental and theoretical efforts have concentrated on the region around \ce{^{100}Sn}{}~\cite{faestermann_2013}, the heaviest known self-conjugate and doubly magic nucleus (${N=Z=50}$), including decay spectroscopy~\cite{2010_Darby_nucleonic_pairing, Lorusso_2012, 2012_Hinke_100Sn_Beta_decay, Park_2018,Auranen_2018, 2019_Lubos_100Sn_beta_decay, 2019_Park, 2020_Park_101Sn_Spectroscopy}, laser spectroscopy~\cite{2022_Vernon_In_moments, Yang_2022, (deG23)}, Coulomb excitation studies~\cite{Vaman_2007,Bader_2013, Guastella_2013}, and mass measurements~\cite{2019_Xu_n_deficient_indium, Hornung2020, 2021_Mougeot_In, Xing2023ndeftin}.
The similar valence orbitals that the protons and neutrons occupy are expected to enhance the effect of proton-neutron pairing, while the proximity of the double shell closure and proton drip line make it a unique laboratory to test our understanding of the strong interaction. 
However, core-excitation effects, i.e. the promotion of nucleons across shell gaps, can complicate the single- or few-particle picture even near shell closures, making accurate theoretical predictions difficult. 

Theoretical approaches to calculate the properties of neutron-deficient nuclei near \ce{^{100}Sn} are computationally costly due to the large configuration space required. 
Nevertheless, the large-scale shell model (LSSM), the Monte Carlo Shell Model, and \mbox{\textit{ab initio}} approaches have been successfully used in the tin region to describe, e.g., $\beta$-decay rates, quadrupole collectivity, and the enhanced magicity in \ce{^{132}Sn}~\cite{2012_Hinke_100Sn_Beta_decay, Coraggio_2015, 2018_Togashi_MC_SM, 2018_Morris_Ab_Initio_100Sn, 2019_Gysbers_Ab_Initio_beta_decay_rates, Zuker_2021}.   

In the indium isotopic chain, the single proton hole below the ${Z=50}$ shell closure provides insight into the effective proton-neutron interaction.
Mass measurements of the ground states in \ce{^{99}In} and \ce{^{100}In} were recently used to test \textit{ab initio} calculations extended to a medium-mass odd-$Z$ isotopic chain~\cite{2021_Mougeot_In}, thus providing valuable input for shell-model coupled-cluster (CCSM) calculations~\cite{2021_Sun_SM_at_100Sn}.
Moreover, recent results from laser spectroscopy revealed the emergence of nuclear collectivity in neutron-rich indium isotopes, with the 9/2$^+$ ground state abruptly departing the single-particle limit below ${N=82}$~\cite{2022_Vernon_In_moments}. 

Nuclear isomers are particularly important for nuclear-structure studies~\cite{jain2021nuclear} and their long lifetimes allow access to a broader range of experimental techniques.
Measurements on the ${N = 50}$ isomer are an important milestone because they will reveal the effects of completely removing neutron excitations from the valence space, especially compared to ${N=82}$.
Its excitation energy will provide direct access to the energy difference between the configurations in which the proton hole occupies the $\pi g_{9/2}$ orbital (ground state) and $\pi p_{1/2}$ orbital (isomer). 

In this Letter, we present measurements of the isomeric excitation energies in neutron-deficient indium isotopes, including the first determination of the excitation energy of $^{99}$In$^{m}$ at the $N=50$ shell closure.
The experimental results are compared to state-of-the-art LSSM~\cite{RMP} and density functional theory (DFT)~\cite{EDF_book} calculations, as well as to \textit{ab initio} calculations using the valence-space in-medium similarity renormalization group (VS-ISMRG)~\cite{Stroberg:2016ung,Stroberg2019} and the CCSM method~\cite{Sun2018}.
Advances in these methods are not only of interest for nuclear shell structure investigations but are also frequently used in metrology, atomic physics, and quantum chemistry~\cite{sayan2020electronprotonratio, Schuessler2020, bartlett2007quantumchemistry}.


The neutron-deficient indium isotopes were produced at the ISOLDE radioactive ion beam facility at CERN~\cite{Borge_2017, 2017_Catherall_ISOLDE} by impinging a $\SI{1.4}{\giga\electronvolt}$ proton beam onto a thick lanthanum carbide target. 
Elements produced by fission, spallation, and fragmentation diffused out of the heated target into a hot tantalum tube, where they were ionized by the hot surface and through an element-selective two-step laser scheme provided by ISOLDE-RILIS~\cite{2017_Fedosseev_RILIS}.  
The radioactive ion beam was then extracted at \SI{30}{\kilo\electronvolt}, mass separated, and delivered to the ISOLTRAP experiment~\cite{2008_Mukherjee}.
There it was cooled and bunched in a linear radio-frequency quadrupole cooler/buncher \mbox{(RFQ-cb)~\cite{HERFURTH2001254}}. 
The bunched beam was then sent at $\SI{3.2}{\kilo\electronvolt}$ to the multi-reflection time-of-flight mass spectrometer (MR-ToF MS)~\cite{2013_wolf}.
After capturing the bunches using the in-trap lift technique and storage of a few tens of milliseconds, the beam was ejected~\cite{2017_Wienholtz} and analyzed by single-ion counting with a time-of-flight detector.
For calibration and optimization, \ce{^{85}Rb+} and \ce{^{133}Cs+} ions from an offline source were used.

The mass $m$ of the ion of interest is extracted from its measured time of flight $t$, compared to two reference masses $m_1$ and $m_2$ with flight times $t_1$ and $t_2$, respectively: 
${\sqrt{m}=C_{\text{ToF}}\Delta_{\text{Ref}}+\Sigma_{\text{Ref}}/2,}$ where ${\Delta_{\text{Ref}}=\sqrt{m_1}-\sqrt{m_2}}$, ${\Sigma_{\text{Ref}}=\sqrt{m_1}+\sqrt{m_2}}$ and ${C_{\text{ToF}}=(2t-t_1-t_2)/\left[2(t_1-t_2)\right]}$~\cite{2013_Wienholtz}.
The excitation energy of an isomeric state ${E=\left[(\Delta t/t_0)^2+2\Delta t/t_0\right]m_0c^2}$ can be directly related to the ToF difference $\Delta t$ with respect to its ground state of mass $m_0$ and ToF $t_0$, with $c$ being the speed of light. 

\begin{table*}[t]

	\centering
	\setlength{\tabcolsep}{6pt}
	\caption[Mass results]{\scriptsize Mass-measurement results for the indium isotopes given as $C_\text{ToF}$ values (with mass excess calculated) for the ground states and as ToF difference $\Delta t$ to the reference mass (with excitation energy calculated) for the isomeric states. Spin assignments $J^\pi$, half-lifes, and reference masses are taken from the \texttt{AME2020} \cite{AME2020} while the literature values marked with an asterisk are taken from Mougeot \textit{et al.} \cite{2021_Mougeot_In}. Values marked with $\#$ are extrapolated or assigned from systematics. The uncertainties given for the mass excesses and the excitation energies correspond to statistical, followed by systematic uncertainties.}

    \scriptsize
	\begin{tabularx}{\textwidth}{l c c c l l l} \hline\hline 
	    & & & & & \multicolumn{2}{c}{Mass excess or exc. energy ($\si{\kilo\electronvolt}$)} \\
		A & J$^\pi$ & Half-life & Ref. ions & $C_{\text{ToF}}$ or $\Delta t$ (ns) & This Letter & Literature \\ \hline 

		99 & $9/2^{+}$\# & $\SI{3.11\pm0.06}{\second}$ & \ce{^{80}Sr^{19}F+}, \ce{^{133}Cs+} & $0.499646429(355)_{\text{stat}}(270)_{\text{syst}}$ & $-61431(12)_{\text{stat}}(8)_{\text{syst}}$ & $-61429(77)^*$\\

		& $1/2^{-}$\# & $\SI{1}{\second}$\# & \ce{^{99gs}In+} & $174(9)_{\text{stat}}(4)_{\text{syst}}$ & $\phantom{-}671(33)_{\text{stat}}(16)_{\text{syst}}$& $\phantom{-}400\#(150\#)$\\

		100 & $6^{+}$\# & $\SI{5.62\pm0.06}{\second}$ & \ce{^{81}Sr^{19}F+}, \ce{^{85}Rb+} & $0.499690777(350)_{\text{stat}}(156)_{\text{syst}}$ & $-64191(11)_{\text{stat}}(5)_{\text{syst}}$ & $\SI{-64178.1\pm2.2}{}^*$\\

		101 & $9/2^{+}$\# & $\SI{15.1\pm1.1}{\second}$ & \ce{^{82}Sr^{19}F+}, \ce{^{133}Cs+} & $0.499677661(69)_{\text{stat}}(99)_{\text{syst}}$ & $-68552.6(93)_{\text{stat}}(28)_{\text{syst}}$ & $-68545.4(47)^*$\\

		& $1/2^{-}$\# & $\SI{10}{\second}$\# & \ce{^{101gs}In+} & $169.3(35)_{\text{stat}}(17)_{\text{syst}}$ & $\phantom{-}658(14)_{\text{stat}}(7)_{\text{syst}}$& $\phantom{-}\SI{668\pm10.8}{}^*$\\
		\hline
	\end{tabularx}
	\label{tab:results}
    
\end{table*}
To achieve the resolving power ${R=t_0/(2\Delta t_\text{FWHM})}$ necessary to separate the indium isomers, the capabilities of the MR-ToF MS were greatly enhanced.
An extended active and passive multi-mirror voltage stabilization system based on Refs.~\cite{2019_Wienholtz, Fischer_2021_voltages} was implemented, which reduced $\Delta t_\text{FWHM}$.
This not only allowed stable continuous operation of the device for more than $70$ hours but also a much higher number of revolutions, between 1500 and 3000, increasing $t_0$.
Furthermore, the initial ion-bunch emittance was optimized by synchronizing the experimental cycle to the $\SI{50}{\hertz}$ AC power line and by fine-tuning the RFQ-cb ejection with respect to its radio-frequency field.

\begin{figure}[t]
    \includegraphics[width=\columnwidth]{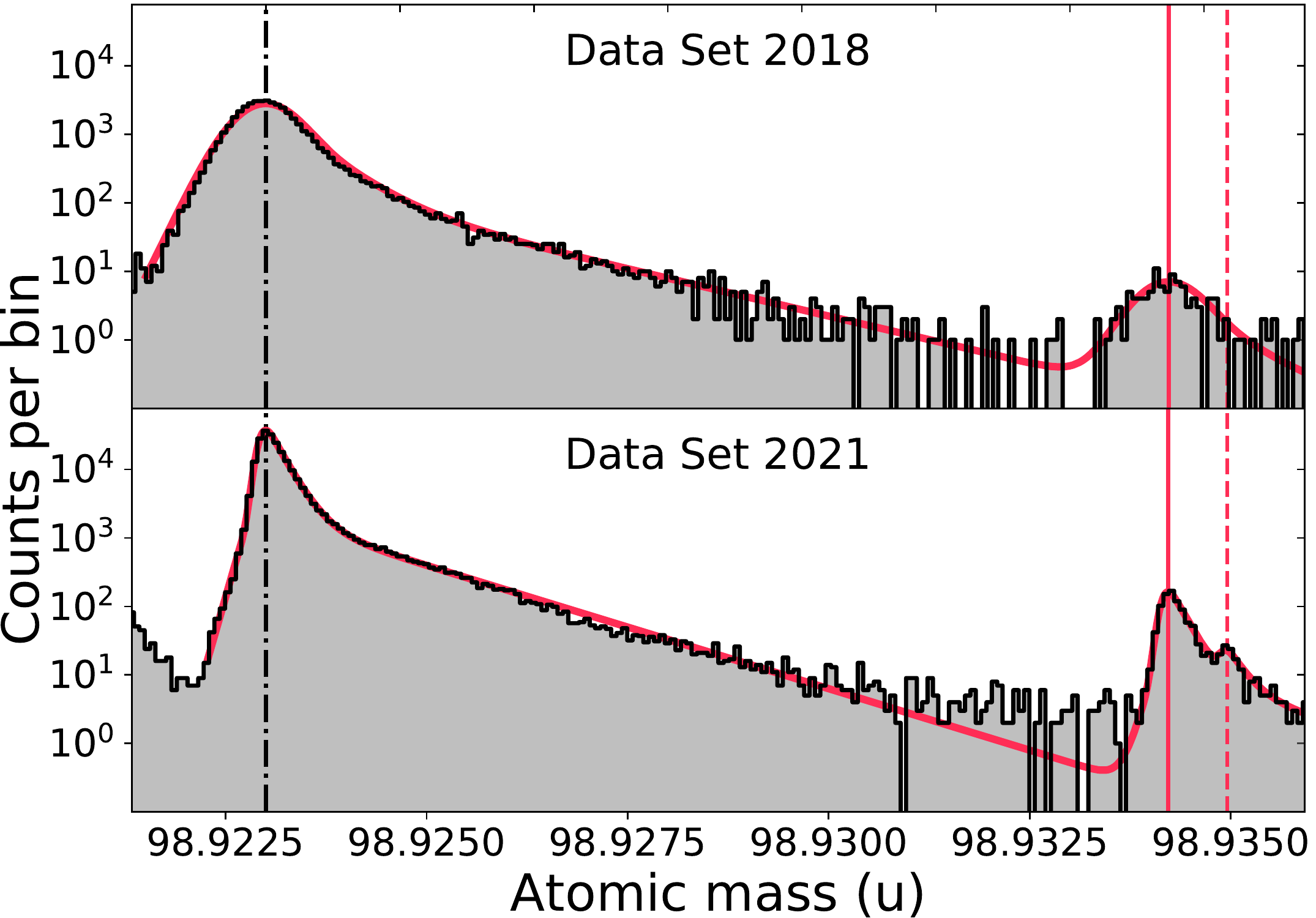}
    \caption{
    Time-of-flight spectrum for the $m/q=99$ beam in the MR-ToF MS with a hyperEMG fit~\cite{2017_Purushothaman_hyperEMG} (red) to the data (see text for explanation).
    The top panel shows the 2018 data set before the device improvements, and the lower panel shows the data set from this paper.
    The black (dash-dotted) line highlights the ToF for the strontium fluoride molecule.
    The vertical red lines show the ground state ToF (solid) and isomeric state ToF (dashed) of \ce{^{99}In}. The bin size is $\sim \SI{73}{\micro\atomicmassunit}$.\label{fig:experimental_results}}
\end{figure}

Figure \ref{fig:experimental_results} shows the ToF spectrum for the ISOLDE beam with mass-to-charge ratio $m/q=99$ compared with the data set from Ref.~\cite{2021_Mougeot_In} to highlight the performance improvement. 
Surface-ionized contamination (here \ce{^{80}Sr^{19}F+}) was identified by calculating its mass $m$ from its observed flight time $t$ and comparing it to the known values of potential isotopes and molecules in the mass region.
Indium was identified by its ToF and RILIS laser on-off tests.
With an average proton current of $\SI{2.0}{\micro\ampere}$ and 
about 3$\times10^{13}$ protons per pulse, roughly four \ce{^{99}In+} ions per second were extracted from the target on average.
The ground-state-to-isomer ratio was determined to be 13:1, resulting in less than $0.3$ isomers per second delivered to the spectrometer.

ToF drifts were eliminated by calculating time-rolling averages of the reference \ce{SrF+} molecule, thus quantifying the drifts and allowing to correct the ToF spectrum during the experiment, similar to Ref.~\cite{2018_Fischer_ToF_Correction}.
For the ToF of $\SI{50}{\milli\second}$ thus obtained for the indium ions, the resulting ToF widths of ${\Delta t_\text{FWHM}\approx\SI{50}{\nano\second}}$ allowed a mass resolving power of ${5\times10^5}$, an improvement factor of 2.5 compared to our previous experiment~\cite{2021_Mougeot_In}.
The improvement helps not only the direct measurement of nuclear isomers and the isobaric purification for Penning-trap measurements but also increases long-term operation stability.

To extract the ground-state $C_\text{ToF}$ value and the excitation energy of the isomeric state, a simultaneous fit of \ce{^{80}Sr^{19}F+} and both \ce{^{99}In+} states was performed.
Due to the asymmetric nature of the ToF distribution, a multi-component exponentially modified Gaussian probability density function (``hyperEMG")~\cite{2017_Purushothaman_hyperEMG} was used.
This approach captures most of the tailing towards longer ToF, while small deviations from the model in the tail showed no influence on the mean of the Gaussian contribution to the fit, i.e. the extracted absolute ToF values. 

To study systematic effects on the data evaluation method, radioactive ion beams were taken for ${99\leq m/q \leq 101}$. 
The results are listed in Table \ref{tab:results}.
The contaminant SrF$^+$ served as the first reference to determine the $C_\text{ToF}$ values, while \ce{^{133}Cs+} from an offline ion source was used as the second reference.
The relative production rates of the two indium states were similar along the investigated chain, suggesting, in combination with laser spectroscopy data~\cite{EBERZ19879}, a $9/2^+$ and $1/2^-$ spin assignment to the ground and isomeric states, respectively.
This is furthermore supported by a recent gamma-spectroscopy experiment of $^{99}$In~\cite{2020_Park_101Sn_Spectroscopy}, which assigns spin $9/2^+$ to the dominantly produced state.
The mass measurement results are in excellent agreement with previous studies \cite{2019_Xu_n_deficient_indium, Hornung2020, 2021_Mougeot_In, Xing2023ndeftin}, improving the precision of our former $^{99}$In ground-state mass measurement by a factor of five. 
Notably, the enhanced MR-ToF MS now achieved a similar precision as the Penning trap experiment from~\cite{2021_Mougeot_In}. 

\begin{figure}[t]
    \subfigure{
       \includegraphics[width=0.99\columnwidth]{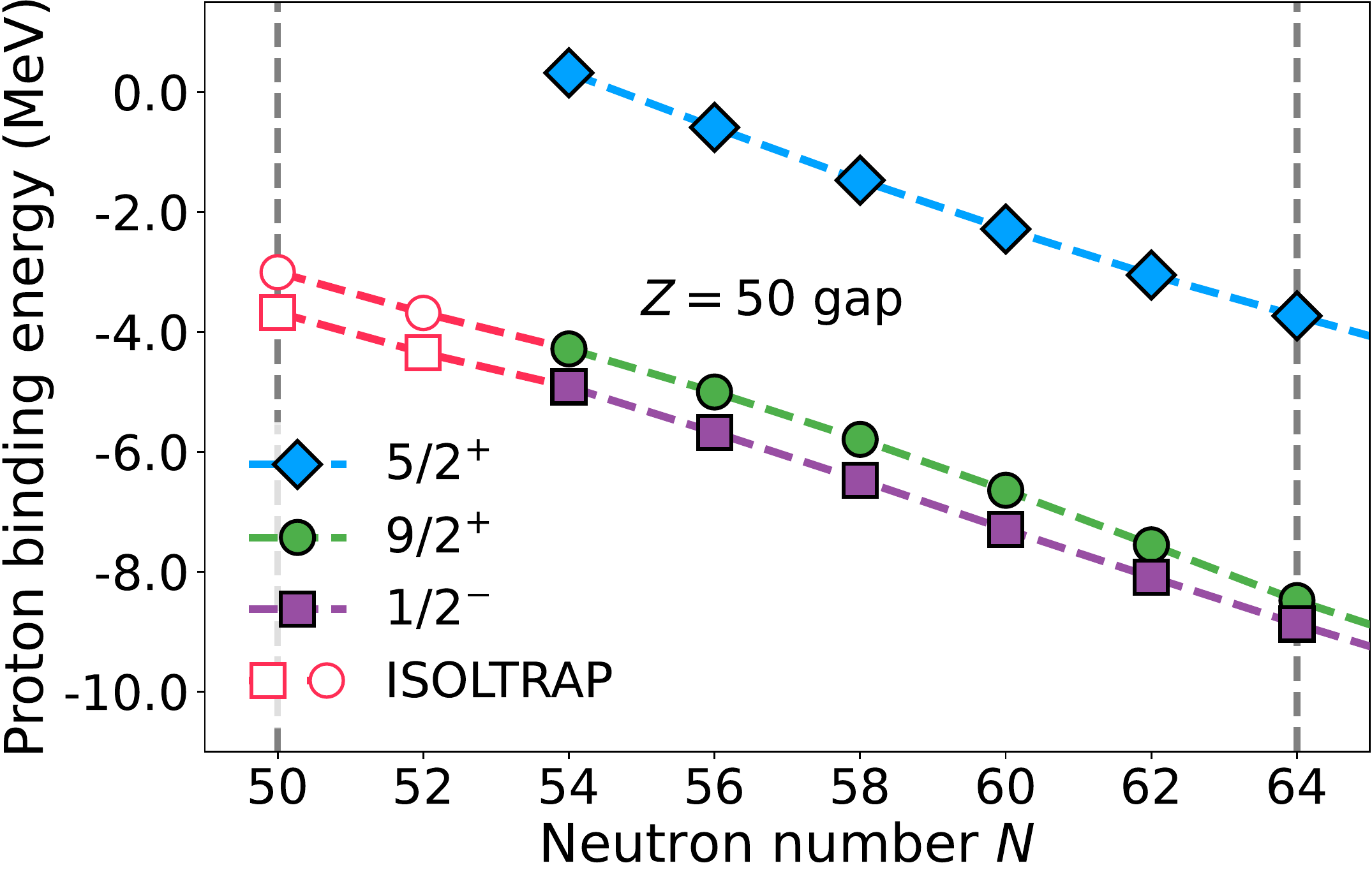}
    }
	\caption{
	 Proton binding energies for nuclear states of the indium ($Z=49$, green and purple) and antimony ($Z=51$, blue) isotopic chains. Data taken from Ref.~\cite{AME2020} (solid symbols) and this work (open symbols, red). The vertical dashed lines indicate the spherical shell closure at $N=50$ and the sub-shell closure at $N=64$.}
	 \label{fig:exp_results}
\end{figure}

In the simplest shell-model picture the ground and isomeric states are formed by a proton hole in the $\pi g_{9/2}$ and the $\pi p_{1/2}$ shells, respectively, determining the spins and parities of the two states.
The evolution of their binding energies with neutron number, presented in Fig.~\ref{fig:exp_results}, is influenced by the filling of the $\nu d_{5/2}$ and $\nu g_{7/2}$ neutron shells. 
The ${Z=50}$ shell gap is formed between the $9/2^+$ states in indium and the $5/2^+$ states in antimony, also shown in Fig.~\ref{fig:exp_results}.
The present measurements extend the data down to ${N=50}$.
Although the experimental binding energies of the two states are not linear with neutron number, the splitting between the two states is almost constant (including that of $^{99}$In, determined in this work to be about $\SI{670}{\kilo\electronvolt}$), only changing at the ${N=64}$ sub-shell closure. 
This is intriguing, considering the large variation in neutron number between ${N=50}$ and ${N=64}$.

\begin{figure}[t]
    \subfigure{
      \includegraphics[width=0.99\columnwidth]{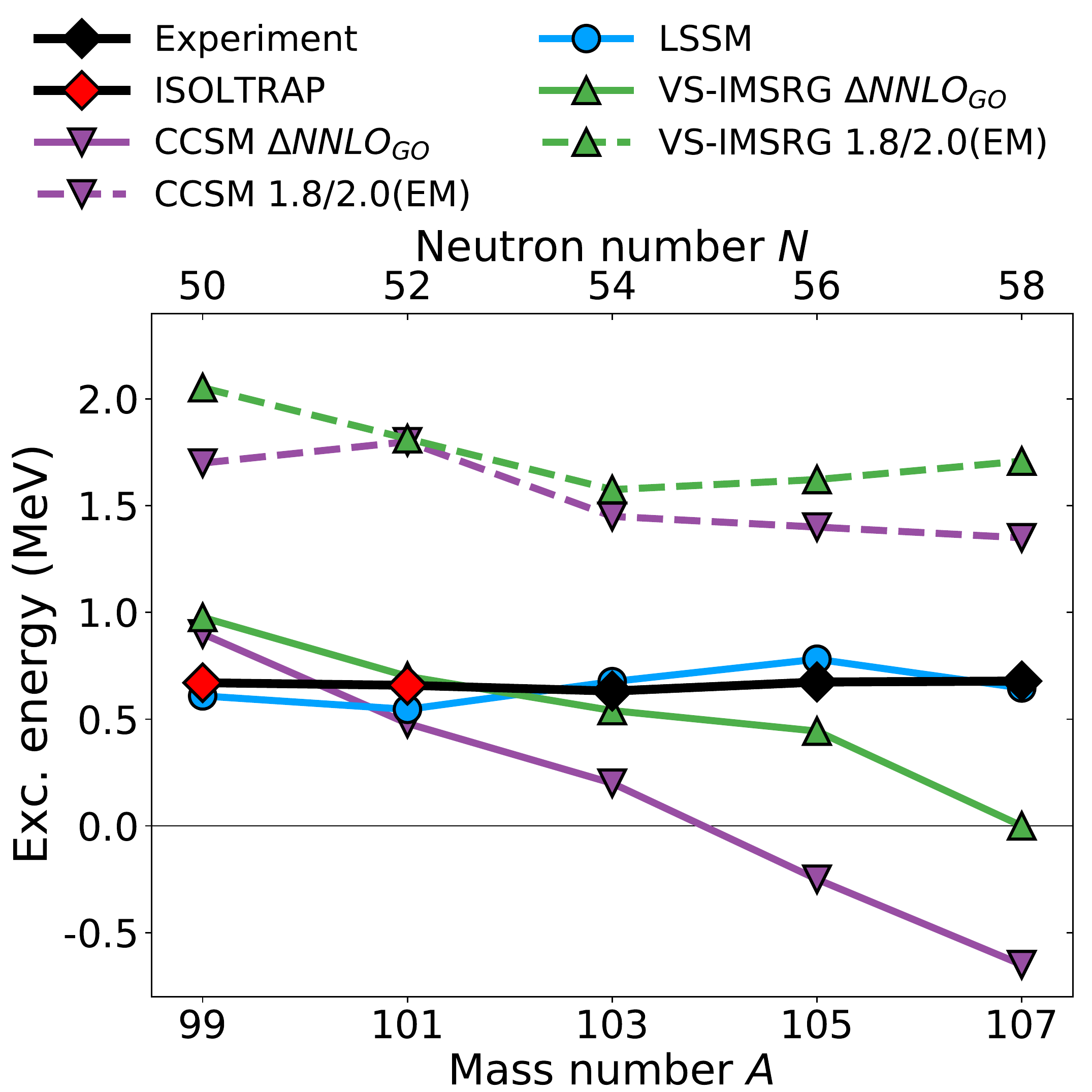}
    }
	\caption{Excitation energies for the $1/2^-$ states in odd-even neutron-deficient indium isotopes compared to large-scale shell model and \mbox{\textit{ab initio}} calculations. The CCSM results are taken from~\cite{2021_Sun_SM_at_100Sn}.}
	\label{fig:exc_energies}
\end{figure}

To investigate the origin of this constant trend, we performed LSSM calculations with the effective interaction above a \ce{^{88}Sr} core employed previously to obtain $\beta$-decay half-lives around ${N=82}$~\cite{2010_Naqvi, 2013_Zhi}. 
To study neutron-deficient indium isotopes, the single-particle energies were adjusted to reproduce the spectrum of $^{91}$Zr, while the $V_{g_{9/2}-g_{7/2}}$ $T=0$ proton-neutron monopole interaction was made more attractive (by $\SI{-600}{\kilo\electronvolt}$) to match the observed shell evolution between $^{91}$Zr and $^{101}$Sn.
The calculations were performed using the Strasbourg shell-model codes Antoine and NATHAN~\cite{ANTOINE, RMP}, maximally allowing for 4-particle--4-hole excitations for both neutrons and protons across the ${N=Z=50}$ gap (3-particle--3-hole for ${A=105,107})$. 
The excitation energies are converged within $\SI{50}{\kilo\electronvolt}$ in all nuclei.
The resulting energy splitting, shown in blue in Fig.~\ref{fig:exc_energies}, is very close to the experimental results. 

The LSSM predicts states with very similar proton orbital occupation across the indium chain.
The neutrons above ${N=50}$ are located predominantly in the $\nu d_{5/2}$ and $\nu g_{7/2}$ orbitals, having little effect on the proton occupancy. 
The results indicate that the attractive monopole interaction between the $\pi g_{9/2}$ and the $\nu g_{7/2}$ nucleons is roughly compensated by the sum of the likewise attractive $T=0$ $\pi p_{1/2}-\nu d_{5/2}$ and $\pi p_{1/2}-\nu g_{7/2}$ monopoles, which is likely the reason for the similar energy splittings of the two states. 
The slight variations of the $1/2^-$ energy may further relate to modiﬁcations of the relative population of neutron orbitals along the isotopic chain.
This complex picture makes the excitation energy of the $1/2^{-}$ isomeric state an interesting benchmark for \textit{ab initio} methods, which have only recently been applied to this region~\cite{2018_Morris_Ab_Initio_100Sn,2021_Sun_SM_at_100Sn,2021_Mougeot_In}.

Thus, we present \textit{ab initio} calculations using the VS-IMSRG and also compare to the CCSM results of Ref.~\cite{2021_Sun_SM_at_100Sn}, using the 1.8/2.0(EM)~\cite{1820,Simo17SatFinNuc} and $\Delta$NNLO$_{GO}$ interactions~\cite{2020_Jiang}.
The VS-IMSRG calculations are performed in a 15 major-shell harmonic oscillator (HO) space.
For the three-nucleon matrix elements, an additional $E_{\rm 3max}$ truncation is required, defined as the sum of three-body HO principal quantum numbers.
Here we use $E_{\rm 3max}=24$ which is sufficiently large in the $A\sim 100$ region~\cite{Miyagi2022} to achieve converged results.
To explicitly capture the effect of excitations across the ${N=Z=50}$ gap, both the proton and neutron $1p_{1/2}$, $1p_{3/2}$, $1d_{5/2}$, $0g_{7/2}$, and $0g_{9/2}$ spaces were decoupled above a $^{88}$Se core using the multi-shell approach of Ref.~\cite{Miyagi2020}.
While the full valence-space diagonalization is impossible, up to 5-particle--5-hole excitations across the ${N=Z=50}$ gap were included.
We observed that the excitation energies are converged to $\approx$\SI{70}{\kilo\electronvolt} with respect to the particle-hole truncation.

As shown in Fig.~\ref{fig:exc_energies}, the two employed interactions result in similar energy-splitting trends for both the VS-IMSRG and CCSM methods. 
The $\Delta$NNLO$_{GO}$ interaction tends to be more accurate at the expense of a linear decrease of excitation energy with $N$ leading to an inversion, which is at odds with the data.
On the other hand, the 1.8/2.0(EM) interaction reproduces the rather flat trend of the splitting better and does not result in any state inversion, but overpredicts the magnitude of the splitting. 

The state crossing in $^{107}$In from the calculations with  $\Delta$NNLO$_{GO}$ can be understood by comparing the diagonal monopole matrix elements of the $\Delta$NNLO$_{GO}$ and 1.8/2.0 (EM) valence-space interactions.
Similar to what is found in the LSSM calculations, 
the flatness of the excitation energies for the 1.8/2.0 (EM) interaction results from the relevant monopole matrix elements and the balanced occupancy of $\nu g_{7/2}$ and $\nu d_{5/2}$. 
This similarity is reinforced by the observation that both interactions reproduce the energy splitting $\nu g_{7/2}-\nu d_{5/2}$ of the single-neutron state in  $^{101}$Sn~\cite{Seweryniak_2007_101Sn_single_n_states} within $\SI{100}{\kilo\electronvolt}$.
Although the relevant matrix elements are almost the same for 1.8/2.0 (EM) and $\Delta$NNLO$_{GO}$, the weaker $\nu p_{3/2} - \nu g_{7/2}$ monopole repulsion of the latter reduces the mixing between $\nu g_{7/2}$ and $\nu d_{5/2}$, leading to a larger $\nu g_{7/2}-\nu d_{5/2}$ splitting.
In this case, the filled $\nu g_{7/2}$ configuration leads to a linear decrease.

\begin{figure}[t]
    \centering
    \subfigure{
    \includegraphics[width=\columnwidth]{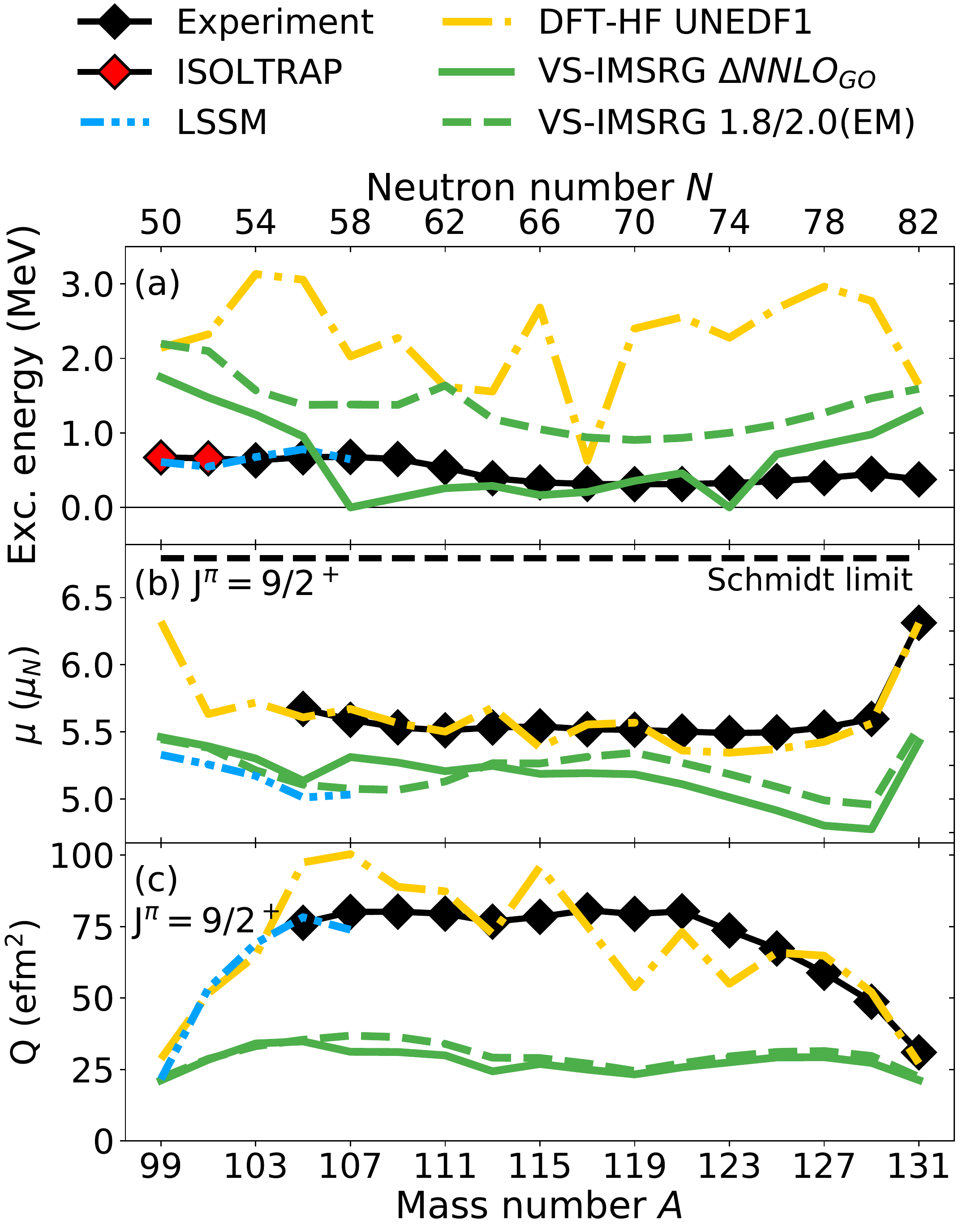}
    }
    \caption{Isomer excitation energies (a) and ground-state electromagnetic moments (b and c) of odd-even indium isotopes, compared to theoretical calculations. The experimental data are from~\cite{ENSDF2022} (black) and this work (red) for the excitation energies and from~\cite{2022_Vernon_In_moments} for the moments. Both DFT-HF and VS-ISMRG calculations are from~\cite{2022_Vernon_In_moments} and are extended in this work to reach ${N=50}$.
    \label{fig:calculations}}
\end{figure}

For a broader view, in Fig.~\ref{fig:calculations} we show the energy splitting across the full indium chain with recent nuclear moment measurements from Ref.~\cite{2022_Vernon_In_moments}, compared to theoretical calculations. 
(Note that for Ref.~\cite{2022_Vernon_In_moments}, the \mbox{VS-IMSRG} calculations were performed in a different valence space than the results shown in Fig.~\ref{fig:exc_energies}, leading to a slightly different energy splitting for $A\leq107$.)  
In addition, we extend the DFT calculations of Ref.~\cite{2022_Vernon_In_moments} to ${N=50}$.
Those were performed within the Hartree-Fock approximation for both protons and neutrons and thus they stagger with $N$ owing to occupying consecutive individual single-particle deformed neutron orbitals.
The complementary data shows another remarkable constancy: that of the magnetic dipole moments of the 9/2$^+$ ground state which are significantly lower than those expected in the single-particle configuration (the so-called Schmidt limit~\cite{Neyens_2003}), except for the value at the ${N=82}$ closed shell.
The excitation energy at ${N=82}$, similar to ${N=50}$, stays rather constant.
From the VS-IMSRG calculations, this can be explained by the monopole strengths, which are almost the same at ${N=50}$ and ${N=82}$, and are only weakly dependent on the number of neutrons.

While the differences between the excitation energy measurements and the DFT calculations are quite large for the density functional UNEDF1~\cite{(Kor12b)} used here, a rather constant trend remains at the same level as the \textit{ab initio} results (see Fig.~\ref{fig:calculations}a).
We note that in the DFT calculations, the isomer excitation energy is directly related to the strength of the spin-orbit interaction (see the discussion of the analogous excitation energy in silver isotopes~\cite{(deG23)}).
Therefore, the measurements presented in this Letter provide an important anchor point for future global readjustments of nuclear density functionals.

By including time-odd fields, the DFT approach accurately reproduces the 9/2$^+$ dipole moments $\mu$.
In contrast, the LSSM and \mbox{VS-IMSRG} calculations underestimate the absolute value but reproduce the general trend well (see Fig.~\ref{fig:calculations}b). 
The sudden increase of the magnetic moment at ${N=82}$ is well described by the DFT calculations, which predict a similar occurrence at ${N=50}$. 
Intriguingly, the LSSM and \textit{ab initio} calculations show a much smoother evolution towards ${N=50}$.
While the excitation energy is flat at ${N=82}$ due to the cancellation of the monopoles in the VS-IMSRG calculations, the 9/2$^+$ dipole moments are much more sensitive to the neutron configuration.   

The DFT and LSSM calculations reproduce the quadrupole moments $Q$ reasonably well, while the VS-IMSRG describes neither the absolute values nor the trend (see Fig.~\ref{fig:calculations}c). 
This is most likely due to collective effects that are not fully captured when calculating the E2 matrix elements at the \mbox{IMSRG(2)} level~\cite{Stroberg2022}.

In summary, the excitation energy of the 1/2$^{-}$ isomer in \ce{^{99}In}{} has been measured for the first time, thanks to significant upgrades of the ISOLTRAP MR-ToF MS at ISOLDE/CERN. 
The systematics of the isomer excitation energy now reach the crucial ${N=50}$ shell closure, confirming its constancy - even with no neutrons left in the valence shell. 
The shell model and the \mbox{{\it ab initio}} calculations using the 1.8/2.0(EM) interaction describe the constancy with the compensation of the monopole proton-neutron interactions, via a balanced occupation of the valence neutron orbitals.
The $\Delta$NNLO$_{GO}$ interaction results in a different occupation and leads to a linear decrease with neutron number, at odds with the experiment.
Examining the electromagnetic moments of the 9/2$^{+}$ ground states and including DFT calculations in the comparisons, we find that all models struggle to describe both energy and electromagnetic observables consistently.
Measurements of nuclear moments of the 1/2$^{-}$ and 9/2$^{+}$ states down to ${N=50}$ are needed to further benchmark the trends predicted by the calculations, as well as future theoretical developments.    

\begin{acknowledgments}
We thank the ISOLDE technical group and the ISOLDE Collaboration for their support. We acknowledge the support of the German Max Planck Society, the French Institut National de Physique Nucléaire et de Physique des Particules (IN2P3), the European Research Council (ERC) under the European Union’s Horizon 2020 research and innovation programme (Grant Agreements No.~682841 ‘ASTRUm’, 654002 ‘ENSAR2’, 101020842 ‘EUSTRONG’, and 861198 ‘LISA’), as well as the German Federal Ministry of Education and Research (BMBF; Grants No.~05P18HGCIA, 05P21HGCI1, and 05P21RDFNB). L.N. acknowledges support from the Wolfgang Gentner Programme of the German Federal Ministry of Education and Research (Grant No.~13E18CHA).
This work was partially supported by the STFC Grant Nos.~ST/P003885/1 and~ST/V001035/1, by the Polish National Science Centre under Contract No.~2018/31/B/ST2/02220, and by a Leverhulme Trust Research Project Grant.
The VS-IMSRG computations were in part performed with an allocation of computing resources at the J\"ulich Supercomputing Center and with an allocation of computing resources on Cedar at WestGrid and Compute Canada using imsrg++~\cite{imsrg++} and KSHELL~\cite{Shimizu2019} codes.
We acknowledge the CSC-IT Center for Science Ltd., Finland, for allocating computational resources.
This project was partly undertaken on the Viking Cluster, which is a high-performance computing facility provided by the University of York. 
We are grateful for computational support from the University of York High Performance Computing service, Viking, and the Research Computing team.

The experiment was conducted by M.A.-K., M.Au, D.A., K.B., I.K., Yu.A.L., D.L., V.M., M.M., L.N., and F.W. The theoretical calculations were performed by J.D., B.S.H, J.D.H, T.M., A.S, and K.S. All authors contributed to the preparation of the manuscript. 
\end{acknowledgments}

\bibliography{bib.bib}

\end{document}